\documentclass[pra,aps,10pt,twocolumn,showpacs,amsmath,amssymb]{revtex4-1}
\usepackage[english]{babel}
\usepackage{afterpage}
\usepackage{bm}
\usepackage{graphicx}
\usepackage{slashed}
\usepackage{latexsym}
\usepackage{amsmath}
\usepackage{amssymb}
\usepackage{color}
\usepackage{lipsum}
\usepackage{float}

\newcommand{\dd}{d}         
\newcommand{\ii}{ {\rm i}}  
\newcommand{\ee}{{\rm e}}  

\begin{document}

\title{Electron supercontinuum in ionization by relativistically intense laser pulses}
\author{F. Cajiao V\'{e}lez}
\author{J. Z. Kami\'{n}ski}
\author{K. Krajewska}
\email[E-mail address:\;]{Katarzyna.Krajewska@fuw.edu.pl}
\affiliation{Institute of Theoretical Physics, Faculty of Physics, University of Warsaw, Pasteura 5,
02-093 Warsaw, Poland}
\date{\today}

\begin{abstract}
Ionization of hydrogen-like ions by intense, circularly polarized laser pulses is analyzed under the scope of the relativistic strong-field approximation. 
We show that, for specific parameters of the laser field, the energy spectra of photoelectrons present a broad region without interference (supercontinuum) which can be controlled by modifying 
the laser field intensity. The physical interpretation of the process is developed according to the Keldysh theory, emphasizing the importance of the complex-time saddle point contributions to the total probability of photoionization. The corresponding polar-angle distributions present an asymmetry attributed to radiation pressure effects. 
\end{abstract}


\maketitle

\section{Introduction}
\label{sec:introduction}

In nonlinear optics, the supercontinuum generation refers to the process for which a narrow-band laser field presents a considerable spectral broadening, resulting in radiation frequencies with a large bandwidth while evincing temporal and/or spatial coherence \cite{bib:dudley,bib:kasia1}. In a similar way, the concept of supercontinuum can be applied to describe the energy spectrum of electrons emitted by photoionization of atoms or ions by strong laser fields. When the energy spectra of photoelectrons exhibit broad structures without important modulations in the scale of tenths (or even hundreds) of laser photon energies, it is considered to present a supercontinuum.

The ionization by strong laser fields is typically dominated by interference effects. The latter are demonstrated, for instance, when the spectrum of photoelectrons consists of series of equally separated peaks. This is also true when the driving pulse comprises just few oscillations. When the monochromatic plane-wave approximation is considered, those peaks are separated by the laser carrier frequency and are commonly recognized as multiphoton peaks. In the pioneering work by Keldysh \cite{bib:keldysh} (see the review by Popov \cite{popov}), it was shown that the total amplitude of photoionization by strong fields contains the contribution of multiple factors arising from different complex-time saddle points. Those contributions lead to pronounced interference effects and to the formation of the aforementioned peaks in the energy spectrum of photoelectrons.

According to the Keldysh theory, if just one complex-time saddle point contributes predominantly to the total probability of photoionization, the interference effects are expected to be suppressed (ionization without interference) leading to the creation of a supercontinuum. On the contrary, if two or more saddle points present similar contributions, interference dominates the process with the consequent formation of series of peaks in the spectra of photoelectrons. 

According to Ref.~\cite{no_interference}, the electron supercontinuum should be observed in photoionization of hydrogen-like ions by short, relativistically intense, and circularly (or elliptically) polarized laser pulses. If the driving field is linearly polarized, the energy spectra of photoelectrons present characteristic interference patterns and the supercontinuum is not formed. Note that, with the current technology, it is possible to obtain laser pulses with intensities larger than $10^{20}$~W/cm$^2$ and durations in the femtosecond regime \cite{tajima,krainov}. Furthermore, the driving field parameters considered here and in Ref.~\cite{no_interference} are experimentally attainable (see, e.g., Refs. \cite{xray1,xray2,xray3}).


The aim of this paper is to investigate the interference-free ionization process for the relativistically intense and short laser pulses.

In this paper we set $\hbar=1$. In our numerical calculations we use relativistic units with $\hbar=m_{\rm e}=c=1$, where $m_{\rm e}$ is the electron rest mass and $c$ is the speed of light. 
Moreover, the relativistic scalar product between two arbitrary four-vectors $a$ and $b$ is written as $a\cdot b=a^\mu b_\mu=a^0b^0-a^1b^1-a^2b^2-a^3b^3$ and the Feynman notation 
$\slashed{a} = \gamma\cdot a=\gamma^{\mu} a_{\mu}$, where $\gamma^{\mu}$ are the Dirac gamma matrices, is used. As usual, $\bar{u}=u^{\dagger}\gamma^0$. When necessary, we employ the 
so-called light-cone variables, i.e., for an arbitrary unit vector ${\bm n}$ and an arbitrary four-vector $a$, we define $a^{\|}=\bm{n}\cdot\bm{a}$, $a^-=a^0-a^{\|}$, $a^+=(a^0+a^{\|})/2$, and $\bm{a}^{\bot}=\bm{a}-a^{\|}\bm{n}$. 
Thus, $a\cdot b=a^+b^-+a^-b^+-\bm{a}^{\bot}\cdot\bm{b}^{\bot}$ and $\dd^4x=\dd x^+\dd x^-\dd^2x^{\bot}$.

\section{Theory}

Let us consider the interaction between a relativistically-intense laser field and a hydrogen-like ion. The exact probability amplitude of ionization is (see, Ref. \cite{no_interference})
\begin{equation}
\mathcal{A}_{\mathrm{fi}}=-\ii \int\dd^4x \ee^{-\ii (E_0/c)x^0}\bar{\Psi}_{\mathrm{f}}(x)e\slashed{A}_{\mathrm{R}}(x)\Psi_{\mathrm{i}}(\bm{x}),
\label{dirac24}
\end{equation}
where the bispinor $\Psi_{\mathrm{i}}(\bm{x})$ describes the electron bound state of energy $E_0$, $\Psi_{\mathrm{f}}(x)$ represents the exact scattering state, $A_{\mathrm{R}}^{\nu}(x)$ is the four-vector potential describing the laser pulse, and $e<0$ is the electron charge.

While the ground state wavefunction for hydrogen-like systems is known exactly (see, e.g., Ref.~\cite{bjorken}), the scattering state $\Psi_{\mathrm{f}}(x)$ has to be approximated, specially when high intensity laser fields are involved. Assuming that the kinetic energy of the photoelectron characterized by the asymptotic momentum ${\bm p}$ is much larger than the ionization potential, i.e., $E_{\bm p}=\sqrt{(m_{\mathrm{e}}c^2)^2+(c\bm{p})^2}-m_{\mathrm{e}}c^2\gg m_\mathrm{e}c^2-E_0$, the scattering state can be approximated by means of the Born expansion. In the zeroth order, such approximation consists in replacing the exact state ${\Psi}_{\mathrm{f}}(x)$ in Eq.~\eqref{dirac24} by the solution of the Dirac equation in the laser field ${\Psi}^{(0)}_{\bm{p}\lambda}(x)$, which does not account for the interaction with the atomic potential,
\begin{equation}
\bigl(\ii\slashed{\partial}-e\slashed{A}_{\mathrm{R}}(x)-m_{\mathrm{e}}c \bigr){\Psi}^{(0)}_{\bm{p}\lambda}(x)=0.
\label{dirac25}
\end{equation}
Here, the subscript $\lambda=\pm$ stands for the electron spin polarization. The solutions of \eqref{dirac25} are known as the Volkov solutions and can be derived exactly for laser fields in the plane-wave front approximation. This is the essence of the relativistic strong-field approximation (RSFA). Therefore, the probability amplitude of ionization under the RSFA [now denoted as $\mathcal{A}(\bm{p},\lambda;\lambda_{\mathrm{i}})$], takes the form
\begin{align}
\mathcal{A}(\bm{p},\lambda;\lambda_{\mathrm{i}})=&-\ii \int\frac{\dd^3q}{(2\pi)^3}\int\dd^4x\, \ee^{-\ii q\cdot x}\bar{\Psi}^{(0)}_{\bm{p}\lambda}(x)\nonumber\\
\times&e\slashed{A}_{\mathrm{R}}(x)\tilde{\Psi}_{\mathrm{i}}(\bm{q}),
\label{dirac26}
\end{align}
where $\tilde{\Psi}_{\mathrm{i}}(\bm{q})$ represents the Fourier transform of the atomic bound state ${\Psi}_{\mathrm{i}}(\bm{x})$. 
In Eq.~\eqref{dirac26} we have introduced $q=(q^0,\bm{q})=(E_0/c,\bm{q})$, which is not a four-vector as it does not transform properly under the relativistic Lorentz transformations. 
Nevertheless, this notation helps us to simplify the formulas presented below. Note that the probability amplitude of ionization $\mathcal{A}(\bm{p},\lambda;\lambda_{\mathrm{i}})$ depends on the initial and final spin states $\lambda_{\mathrm{i}}$ and $\lambda$, respectively, which are denoted as '$+$' for a spin up and '$-$' for a spin down.

Up to now, our considerations have been very general. In the remaining part of this paper the calculations are going to be carried out in the velocity gauge. In order to proceed, we model the electromagnetic potential describing the laser field using the plane-wave front approximation,
\begin{equation}
A_{\mathrm{R}}(x)\equiv A(\phi)=A_0[\varepsilon_1 f_1(\phi)+\varepsilon_2 f_2(\phi)],
\label{dirac27}
\end{equation}
where $\phi=k\cdot x=k^0x^-$, $k=k^0n=k^0(1,\bm{n})$, $k^0=\omega/c$, and $\omega=2\pi/T_{\mathrm{p}}$. In our notation $\omega$ represents the fundamental frequency of the pulse and $T_{\mathrm{p}}$ corresponds to its duration. The polarization of the laser field is determined by two real and normalized four-vectors, $\varepsilon_j\equiv (0,{\bm{\varepsilon}}_j)$, which are perpendicular to the pulse propagation direction (i.e., $k\cdot\varepsilon_j=-{\bm k}\cdot{\bm \varepsilon}_j=0$). The two shape functions, $f_j(\phi)$, are real, with continuous second derivatives, and vanish for $\phi<0$ and $\phi>2\pi$. The unitary vector $\bm{n}$ represents the direction of propagation of the laser pulse.

The Volkov solution for the vector potential~\eqref{dirac27} is given by \cite{no_interference,recoil}
\begin{align}
\psi^{(+)}_{\bm{p}\lambda}(x)=&\sqrt{\frac{m_{\mathrm{e}}c^2}{VE_{\bm{p}}}}\Bigl(1+\frac{m_{\mathrm{e}}c\mu}{2p\cdot k}\bigl[f_1(k\cdot x)\slashed{\varepsilon}_1\slashed{k}\nonumber\\
+& f_2(k\cdot x)\slashed{\varepsilon}_2\slashed{k} \bigr] \Bigr) \ee^{-\ii S_p^{(+)}(x)}u^{(+)}_{\bm{p}\lambda},
\label{dirac31}
\end{align}
where
\begin{align}
S_p^{(+)}(x)&= p\cdot x+\int_0^{k\cdot x}\dd\phi\Bigl[-\frac{m_{\mathrm{e}}c\mu}{p\cdot k}\bigl(\varepsilon_1\cdot p f_1(\phi)\nonumber\\
&+ \varepsilon_2\cdot p f_2(\phi)\bigr)+\frac{(m_{\mathrm{e}}c\mu)^2}{2p\cdot k}\bigl(f_1^2(\phi)+f_2^2(\phi)\bigr)\Bigr].
\label{dirac32}
\end{align}
In Eqs.~\eqref{dirac31} and \eqref{dirac32}, the superscript $(+)$ indicates that $\psi^{(+)}_{\bm{p}\lambda}(x)$ is a positive-energy solution of the Dirac equation in the laser field [Eq.~\eqref{dirac25}], $V$ represents the quantization volume, and $p=(p^0,\bm{p})=(E_{\bm{p}}/c,\bm{p})$ is the on-mass-shell four-vector. In addition, the Dirac free particle bispinors $u^{(+)}_{\bm{p}\lambda}$ are normalized such that $\bar{u}^{(+)}_{\bm{p}\lambda}u^{(+)}_{\bm{p}\lambda'}=\delta_{\lambda\lambda'}$. Note that we have introduced the dimensionless relativistically invariant parameter $\mu$, defined as
\begin{equation}
\mu=\frac{|e|A_0}{m_{\mathrm{e}}c}.
\label{dirac28}
\end{equation}
This parameter is related to the relativistic character of ionization. If $\mu\ll1$ then the quantum mechanical evolution of the field-particle interaction can be analyzed according to the Schr\"{o}dinger equation. In contrast, if $\mu\approx1$ or $\mu>1$, which is the case studied in this paper, relativistic effects need to be accounted for and the time-evolution of the system needs to be studied according to the Dirac equation.

As it is shown in Ref.~\cite{no_interference}, the probability amplitude of ionization [Eq.~\eqref{dirac26}] for a laser pulse described by Eq.~\eqref{dirac27} can be written, in the velocity gauge, as
\begin{equation}
\mathcal{A}(\bm{p},\lambda;\lambda_{\mathrm{i}})=\int \frac{\dd^3q}{(2\pi)^3}\int\dd^4x \ee^{\ii S_p^{(+)}(x)-\ii q\cdot x} M_{\lambda,\lambda_{\mathrm{i}}}(k\cdot x),
\label{dirac34}
\end{equation}
where
\begin{align}
&M_{\lambda,\lambda_{\mathrm{i}}}(k\cdot x)=\ii m_{\mathrm{e}}c\mu\sqrt{\frac{m_{\mathrm{e}}c^2}{VE_{\bm{p}}}}\nonumber\\
&\times \bigl[f_1(k\cdot x)B^{(1,0)}_{\bm{p}\lambda;\lambda_{\mathrm{i}}}(\bm{q})+f_2(k\cdot x)B^{(0,1)}_{\bm{p}\lambda;\lambda_{\mathrm{i}}}(\bm{q})\nonumber\\
&-\frac{m_{\mathrm{e}}c\mu}{2p\cdot n}\bigl([f_1(k\cdot x)]^2+[f_2(k\cdot x)]^2\bigr)B^{(0,0)}_{\bm{p}\lambda;\lambda_{\mathrm{i}}}(\bm{q})\bigr].
\label{dirac34a}
\end{align}
Here, in order to simplify the notation, we have introduced the following functions related to the Fourier transform of the ground-state wavefunction $\Psi_{\mathrm{i}}(\bm{x})$ and the Dirac free particle bispinors,
\begin{align}
B^{(0,0)}_{\bm{p}\lambda;\lambda_{\mathrm{i}}}(\bm{q})=&\bar{u}^{(+)}_{\bm{p}\lambda}\slashed{n}\tilde{\Psi}_{\mathrm{i}}(\bm{q}),\nonumber\\
B^{(1,0)}_{\bm{p}\lambda;\lambda_{\mathrm{i}}}(\bm{q})=&\bar{u}^{(+)}_{\bm{p}\lambda}\slashed{\varepsilon}_1\tilde{\Psi}_{\mathrm{i}}(\bm{q}),\nonumber\\
B^{(0,1)}_{\bm{p}\lambda;\lambda_{\mathrm{i}}}(\bm{q})=&\bar{u}^{(+)}_{\bm{p}\lambda}\slashed{\varepsilon}_2\tilde{\Psi}_{\mathrm{i}}(\bm{q}). \label{dirac33}
\end{align}
Now, defining the laser-dressed momentum $\bar{p}$ as~\cite{no_interference,KKC,KKBW,KMK2013}
\begin{align}
\bar{p}=&p-\frac{m_{\mathrm{e}}c\mu}{p\cdot k}(\varepsilon_1\cdot p\langle f_1\rangle+\varepsilon_2\cdot p\langle f_2\rangle )k\nonumber\\
+&\frac{(m_{\mathrm{e}}c\mu)^2}{2p\cdot k}(\langle f_1^2\rangle+\langle f_2^2\rangle )k,
\label{dirac35}
\end{align}
the function $S_p^{(+)}(x)$ in Eq.~\eqref{dirac34} is rewritten as
\begin{equation}
S_p^{(+)}(x)=\bar{p}^+x^-+p^-x^+-\bm{p}^{\bot}\cdot\bm{x}^{\bot}+G_p(k^0x^-),
\label{dirac37}
\end{equation}
where
\begin{align}
&G_p(\phi)=\int_0^{\phi}\dd\phi' \Bigl[-\frac{m_{\mathrm{e}}c\mu}{p\cdot k}\bigl(\varepsilon_1\cdot p (f_1(\phi')-\langle f_1\rangle) \nonumber \\
& + \varepsilon_2\cdot p (f_2(\phi')-\langle f_2\rangle)\bigr)+\frac{(m_{\mathrm{e}}c\mu)^2}{2p\cdot k}\bigl(f_1^2(\phi') - \langle f_1^2\rangle\nonumber\\
& + f_2^2(\phi')-\langle f_2^2\rangle\bigr)\Bigr].
\label{dirac38}
\end{align}
Note that in Eqs.~\eqref{dirac35} and \eqref{dirac38} the average of a function $F(\phi)$, which vanishes for $\phi<0$ and $\phi>2\pi$, is given by
\begin{equation}
\langle F\rangle=\frac{1}{2\pi}\int_0^{2\pi}\dd\phi F(\phi).
\label{dirac36}
\end{equation}

According to Ref.~\cite{no_interference}, the multidimensional integral in Eq.~\eqref{dirac34} can be treated analytically with the help of the following Fourier transforms defined for $j=1,2$ and $0\leqslant\phi\leqslant 2\pi$,
\begin{align}
\bigl[f_1(\phi)\bigr]^j\exp[\ii G_p(\phi)]=&\sum_{N=-\infty}^{\infty}G_N^{(j,0)}\ee^{-\ii N\phi},\label{dirac39a} \\
\bigl[f_2(\phi)\bigr]^j\exp[\ii G_p(\phi)]=&\sum_{N=-\infty}^{\infty}G_N^{(0,j)}\ee^{-\ii N\phi}.\label{dirac39}
\end{align}
Hence, the probability amplitude of ionization $\mathcal{A}(\bm{p},\lambda;\lambda_{\mathrm{i}})$ can be represented as
\begin{equation}
\mathcal{A}(\bm{p},\lambda;\lambda_{\mathrm{i}})=\ii m_{\mathrm{e}}c\mu\sqrt{\frac{m_{\mathrm{e}}c^2}{VE_{\bm{p}}}}\mathcal{D}(\bm{p},\lambda;\lambda_{\mathrm{i}}),
\label{dirac40}
\end{equation}
were $\mathcal{D}(\bm{p},\lambda;\lambda_{\mathrm{i}})$ involves an infinite sum,
\begin{align}
\mathcal{D}(\bm{p},\lambda;\lambda_{\mathrm{i}})&=\sum_{N=-\infty}^{\infty}\frac{\ee^{2\pi\ii(\bar{p}^+-q^+-Nk^0)/k^0}-1}{\ii (\bar{p}^+-q^+-Nk^0)} \nonumber\\
&\times\Bigl[G^{(1,0)}_N B^{(1,0)}_{\bm{p}\lambda;\lambda_{\mathrm{i}}}(\bm{Q})+ G^{(0,1)}_N B^{(0,1)}_{\bm{p}\lambda;\lambda_{\mathrm{i}}}(\bm{Q}) \nonumber \\
&-\frac{m_{\mathrm{e}}c\mu}{2p\cdot n}[G^{(2,0)}_N+G^{(0,2)}_N]B^{(0,0)}_{\bm{p}\lambda;\lambda_{\mathrm{i}}}(\bm{Q})\Bigr],\label{dirac41}
\end{align}
and 
\begin{equation}
\bm{Q}=\bm{p}+(q^0-p^0)\bm{n}. \label{q}
\end{equation}

Taking into account Eq.~\eqref{dirac40} and keeping in mind that the final density of electron states, according to our current normalization conventions, is equal to $V\dd^3p/(2\pi)^3$, the spin-dependent probability of ionization is given by
\begin{equation}
P(\lambda;\lambda_{\mathrm{i}})=\mu^2\frac{(m_{\mathrm{e}}c)^3}{(2\pi)^3}\int\frac{\dd^3p}{p^0}|\mathcal{D}(\bm{p},\lambda;\lambda_{\mathrm{i}})|^2.
\label{dirac43}
\end{equation}
On the other hand, the initial-spin-averaged triply-differential probability distribution, which is obtained by averaging the values corresponding to the initial spin states and summing up the values corresponding to the final spin states, in atomic units, takes the form
\begin{align}
\mathcal{P}(\bm{p})&=\frac{\alpha^2m_{\mathrm{e}}c^2}{2}\sum_{\lambda,\lambda_{\rm i}=\pm}\frac{\dd^3P(\bm{p},\lambda;\lambda_{\mathrm{i}})}{\dd E_{\bm{p}}\dd^2\Omega_{\bm{p}}}\nonumber\\
&\equiv\frac{\alpha^2\mu^2}{2}\frac{(m_{\mathrm{e}}c)^4}{(2\pi)^3}\sum_{\lambda,\lambda_{\rm i}=\pm}|\bm{p}|\cdot|\mathcal{D}(\bm{p},\lambda;\lambda_{\mathrm{i}})|^2,
\label{dirac44}
\end{align}
where $\alpha=e^2/(4\pi\varepsilon_0c)$ is the fine-structure constant. 

Up to now we have calculated the probability distribution of photoionization of hydrogen-like systems in the RSFA framework and the velocity gauge. The fact that short laser pulses were considered assures that the electron ground-state wavefunction is well-defined and unperturbed before the interaction begins, even for arbitrarily intense laser fields. In contrast, when the infinite plane-wave approximation is used (see, e.g., Refs.~\cite{eikHeidelberg,Klaiber,eikHeidelberg2}), the analysis is restricted to the interaction of highly charged positive ions with fields of moderate intensity. This is in order to guarantee that the ground-state wavefunction is not heavily distorted by the action of the oscillating laser field.

Even though Eqs.~\eqref{dirac40} and \eqref{dirac41} allow us to calculate the energy and angular probability amplitude of ionization for given initial and final spin states, they do not offer a simple insight into the physics behind the process. For this reason, in the next Section, we introduce the saddle-point analysis of the integrals in Eq.~\eqref{dirac34} with the sole purpose of interpreting our numerical results.

\section{Saddle-point approximation}
\label{saddle_point}
To perform the saddle-point analysis of the integrals in the probability amplitude of ionization, we rewrite Eq.~\eqref{dirac34} in terms of the light-cone variables,

\begin{widetext}
\begin{equation}
\mathcal{A}(\bm{p},\lambda;\lambda_{\mathrm{i}})=\frac{1}{k^0}\int_0^{2\pi}\dd\phi \int\frac{\dd^3q}{(2\pi)^3}\int\dd x^+\dd^2x^{\bot} \ee^{\ii (p^--q^-)x^+ -\ii(\bm{p}^{\bot}-\bm{q}^{\bot})\cdot\bm{x}^{\bot}} \ee^{\ii G(\phi)}M_{\lambda,\lambda_{\mathrm{i}}}(\phi),
\label{saddle1}
\end{equation}
where $M_{\lambda,\lambda_{\mathrm{i}}}(\phi)$ is defined by \eqref{dirac34a} and
\begin{equation}
G(\phi)\equiv G(g_0,g_1,g_2,h;\phi)=\int_0^{\phi}\dd\phi'  \bigl[g_0+g_1f_1(\phi')+g_2f_2(\phi') +h\bigl(f_1^2(\phi')+f_2^2(\phi')\bigr)\bigr].
\label{saddle2}
\end{equation}
\end{widetext}
Here we have introduced the functions
\begin{equation}
g_0=\frac{p^+-q^+}{k^0},  \;  h=\frac{(m_{\mathrm{e}}c\mu)^2}{2k\cdot p}, \; g_j=-m_{\mathrm{e}}c\mu\frac{\varepsilon_j\cdot p}{k\cdot p},
\label{saddle3}
\end{equation}
for $j=1,2$. The integration over $\dd x^+\dd^2x^{\bot}$ leads to the conservation relations 
\begin{equation}
p^-=q^- \quad \textrm{and}\quad \bm{p}^{\bot}=\bm{q}^{\bot},
\label{dirac38a}
\end{equation}
and allows us to perform the integration over $\dd^3q$. Finally, the probability amplitude of ionization takes the form
\begin{equation}
\mathcal{A}(\bm{p},\lambda;\lambda_{\mathrm{i}})=\frac{1}{k^0}\int_0^{2\pi}\dd\phi\,\ee^{\ii G(\phi)}\bigl[M_{\lambda,\lambda_{\mathrm{i}}}(\phi)\bigr]_{\bm{q}=\bm{Q}},
\label{saddle4}
\end{equation}
where $\bm{Q}$ is defined in Eq.~\eqref{q}. Note that, according to the relations \eqref{dirac38a}, $g_0={(p^0-q^0)}/{k^0}$ and it only depends on the energy of the initial and final states.

As the function $\ee^{\ii G(\phi)}$ is considered to be fast oscillating compared to the remaining parts of the integrand in Eq.~\eqref{saddle4}, the standard saddle-point method can be used to approximate this expression. The saddle points are obtained by solving the equation
\begin{equation}
\frac{{\rm d}G(\phi)}{{\rm d}\phi}=0,
\label{saddle6}
\end{equation}
which, in general, has complex solutions. The only saddle points that contribute to the integral, denoted as $\phi_s$, are those which satisfy the relation $\textrm{Im}\,G(\phi_s)>0$. With that in mind, the probability amplitude of ionization is approximated as 
\begin{equation}
\mathcal{A}(\bm{p},\lambda;\lambda_{\mathrm{i}})=\frac{1}{k^0}\sum_s \ee^{\ii G(\phi_s)}\sqrt{\frac{2\pi\ii}{G^{\prime\prime}(\phi_s)}}\bigl[M_{\lambda,\lambda_{\mathrm{i}}}(\phi_s)\bigr]_{\bm{q}=\bm{Q}}.
\label{saddle7}
\end{equation}
By analyzing the previous expression, it is clear that the interference pattern in photoionization arises when two or more saddle points contribute importantly to the probability amplitude $\mathcal{A}(\bm{p},\lambda;\lambda_{\mathrm{i}})$. In contrast, if just one of them is dominant over a range of photon energies and emission angles [i.e., if the $\textrm{Im}\,G(\phi_s)$ is considerable small compared to the corresponding value of the remaining saddle points], then no strong interference effects are expected. Therefore, we anticipate that the supercontinuum should appear in the energy range for which just one saddle point contributes the most to Eq.~\eqref{saddle7}. This was demonstrated for the case of fixed initial and final spin states in Ref.~\cite{no_interference}.

\section{Numerical calculations}
\label{numerical_calculations}
We consider the photoionization of He$^+$ ions (with atomic number $Z=2$) by a relativistically-intense and circularly polarized laser pulse. The latter is characterized by the electric field shape functions with a $\sin^2$ envelope defined as 
\begin{equation}
F_j(\phi)=F_0(\phi,\delta_j,\chi)\cos(\delta+\delta_j),
\label{shapes2}
\end{equation}
with
\begin{equation}
F_0(\phi,\delta_j,\chi)=N_0\sin^2\Bigl(\frac{\phi}{2}\Bigr)\sin(N_{\mathrm{osc}}\phi+\delta_j+\chi)
\label{shapes3}
\end{equation}
for $0<\phi <2\pi$ and 0 otherwise. Here, $N_{\mathrm{osc}}$ represents the number of field oscillations within the pulse, $\chi$ is the carrier-envelope phase, $\delta$ and $\delta_j$ determine the polarization properties of the pulse, and $N_0$ is a normalization constant chosen such that the average intensity of the field is independent of the number of cycles (see, Ref.~\cite{no_interference}). The function $F_0$, explicitly written as a function of time, is
\begin{align}
F_0({\bm r}, t,\delta_j,\chi)=&N_0\sin^2\Bigl(\frac{1}{2N_{\mathrm{osc}}}\omega_{\mathrm{L}} (t-\bm{n}\cdot\bm{r}/c)\Bigr)\nonumber\\
\times& \sin(\omega_{\mathrm{L}} (t-\bm{n}\cdot\bm{r}/c)+\delta_j+\chi),
\label{shapes4ex}
\end{align}
for $0<t-\bm{n}\cdot\bm{r}/c <T_{\mathrm{p}}$ and it is $0$ otherwise. Here we have introduced the carrier frequency of the laser field, $\omega_{\mathrm{L}}=N_{\mathrm{osc}}\omega$.

For the numerical calculations presented below we have chosen a circularly polarized laser pulse ($\delta_1=0$, $\delta_2=\pi/2$, and $\delta=\pi/4$), propagating along the $z$-axis ($\bm{n}={\bm e}_z$), and comprising four field oscillations within the $\sin^2$ envelope ($N_{\mathrm{osc}}=4$). The polarization vectors are ${\bm \varepsilon}_1={\bm e}_x$ and ${\bm \varepsilon}_2={\bm e}_y$. The carrier-envelope phase is $\chi=\pi/2$ and the carrier frequency is $\omega_{\mathrm{L}}=20$eV. 

As the electric field is related to the vector potential by the relation $\bm{\mathcal{E}}(\phi)=-\partial_t\bm{A}(\phi)$, the shape functions $f_1(\phi)$ and $f_2(\phi)$ in Eq.~\eqref{dirac27} are calculated as
\begin{equation}
f_j(\phi)=-\int_0^{\phi}\dd \phi' F_j(\phi'),
\label{shapes4}
\end{equation}
with $j=1,2$. 

\begin{figure*}
\begin{center}
\includegraphics[width=10.7cm]{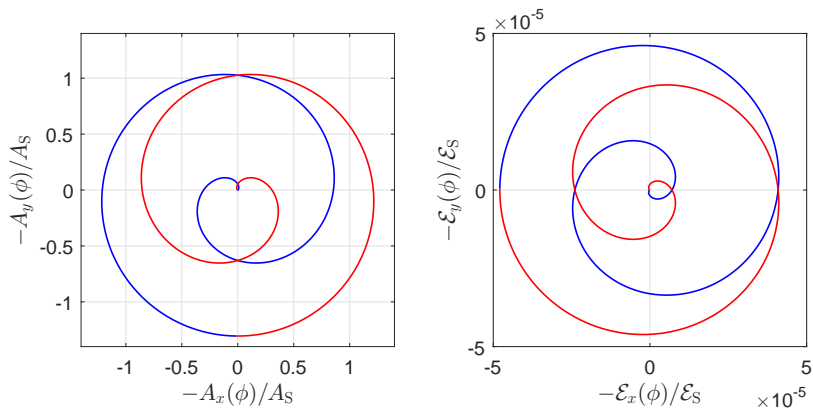}
\end{center}
\caption{(Color online) Trajectories of the tips of the electromagnetic vector potential $\bm{A}(\phi)$ (left panel) and the electric field vector $\bm{\mathcal{E}}(\phi)$ (right panel), 
in relativistic units, for the laser pulse discussed below. All trajectories start from and end up at the origin $(0,0)$. In order to show 
the direction of the time-evolution, we mark with colors blue and red the ramp up and ramp down parts of the laser pulse, respectively. We observe the azimuthal symmetry of the electromagnetic potential, $\varphi\rightarrow \pi-\varphi\textrm{ mod }2\pi$ or $(x,y)\rightarrow (-x,y)$, and of the electric field, $\varphi\rightarrow -\varphi\textrm{ mod }2\pi$ or $(x,y)\rightarrow (x,-y)$. The time-averaged intensity of the laser pulse is $I=4\times 10^{20}$~W/cm$^2$, the carrier laser frequency is $\omega_{\mathrm{L}}=20$~eV, and $N_{\mathrm{osc}}=4$. Note that for these parameters, $\mu >1$.} \label{fig:potentials}
\end{figure*}

In Fig.~\ref{fig:potentials}, we present the time evolution of the tips of the vector potential (left panel) and electric field (right panel) in the $xy$-plane 
(which is perpendicular to the laser field propagation direction) for the pulse described above with an average intensity of $I=4\times 10^{20}$~W/cm$^2$. 
Both curves start at the origin of coordinates and evolve counterclockwise during the ramp up (blue color) and ramp down (red color), and are presented in relativistic units. The parameter $A_{\mathrm{S}}=m_{\mathrm{e}}c/|e|$ and the ratio ${A(\phi)}/{A_{\mathrm{S}}}$ is given by
\begin{equation}
\frac{A(\phi)}{A_{\mathrm{S}}}=\mu[\varepsilon_1 f_1(\phi)+\varepsilon_2 f_2(\phi)].
\label{ratio}
\end{equation}
Note that, for this particular intensity, $|A(\phi)|/A_{\mathrm{S}}$ reaches values larger than the unity, so one needs to consider the problem in a fully relativistic way. On the other hand, the parameter $\mathcal{E}_{\mathrm{S}}=m_{\mathrm{e}}^2c^3/|e|$, known as the Sauter-Schwinger critical electric field (see, Refs.~\cite{no_interference,salamin} and references therein), is related to the probability of electron-positron pair production. As the ratio $|\bm{\mathcal{E}}(\phi)|/\mathcal{E}_{\mathrm{S}}\ll 1$, such effect can be ignored in our calculations.

\subsection{Energy and polar-angle spectra of photoelectrons}
\begin{figure*}
\begin{center}
\includegraphics[width=10.7cm]{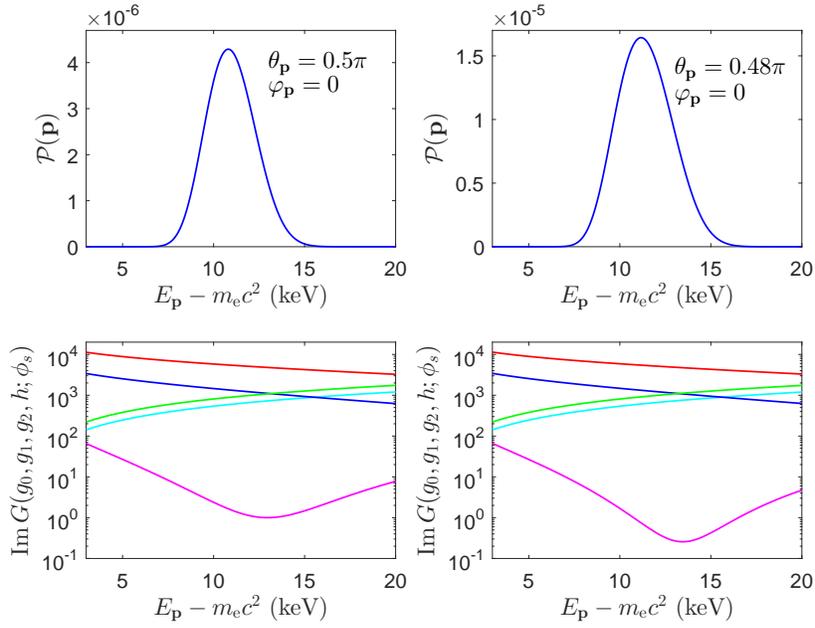}
\end{center}
\caption{(Color online) Initial-spin-averaged energy spectra of photoelectrons (upper row) for the time-averaged intensity $I=2\times 10^{20}$~W/cm$^2$ and for the emission angles indicated in the figure. In the lower row, the appropriate plots of $\textrm{Im}\,G(\phi_s)$ for all five relevant saddle points are shown.} \label{fig:spectra2e20}
\end{figure*}

In the upper panels of Fig.~\ref{fig:spectra2e20} we present the energy spectra of photoelectrons calculated according to Eq.~\eqref{dirac44}. The results are plotted for the electron asymptotic momentum ${\bm p}$ with azimuthal angle $\varphi_{\bm p}=0$ and polar angles $\theta_{\bm p}=0.5\pi$ (left panel) and $\theta_{\bm p}=0.48\pi$ (right panel). While the averaged intensity is $I=2\times 10^{20}$~W/cm$^2$, the remaining parameters are the same as in Fig.~\ref{fig:potentials}. We observe that the energy spectra of photoelectrons present a single and very broad structure which ranges from $3$~keV up to $20$~keV (corresponding to hundreds of single-photons energy), without the distinctive signatures of interference; the supercontinuum. Moreover, the distributions exhibit a maximum located at positions depending on the polar angle. For $\theta_{\bm p}=0.5\pi$, the spectral maximum appears at electron kinetic energies close to $10.96$~keV, whereas for $\theta_{\bm p}=0.48\pi$ it is shifted to $11.17$~keV. 

It follows from our analysis in Sec.~\ref{saddle_point} and the results shown in Ref.~\cite{no_interference} that the presence of a supercontinuum in the energy distributions 
can be related to absence of interference effects, i.e., to the existence of a single saddle point which contributes the most to the probability amplitude of ionization \eqref{saddle7}. 
In the lower panels of Fig.~\ref{fig:spectra2e20} we present the energy dependence of $\textrm{Im}\,G(\phi_s)$ for all relevant saddle points. It can be seen that, for the parameters 
chosen in our calculations, there is just one of such points for which $\textrm{Im}\,G(\phi_s)$ is much smaller than the others (magenta curves in the lower panels). Note that its minimum is located around the region for which the probability distribution is maximal. 

\begin{figure*}
\begin{center}
\includegraphics[width=10.7cm]{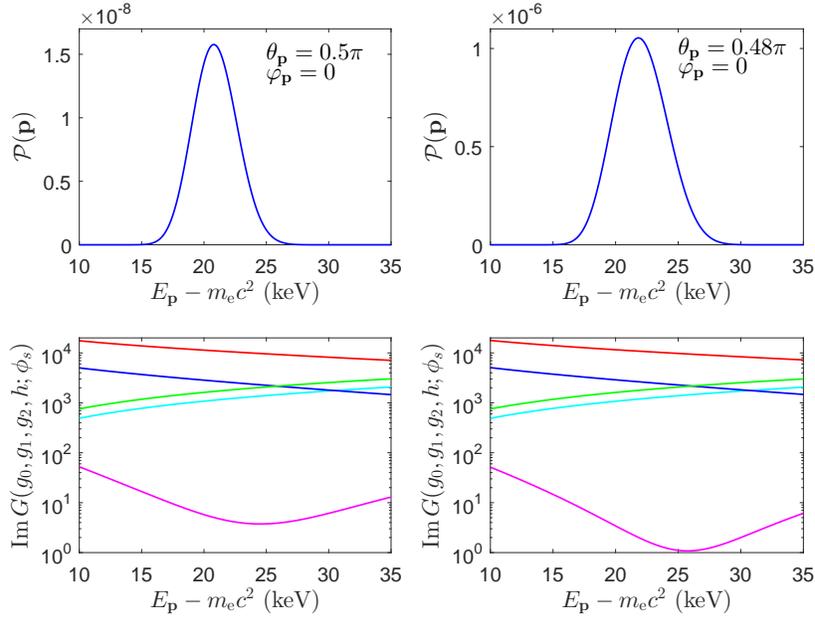}
\end{center}
\caption{(Color online) The same as Fig.~\ref{fig:spectra2e20} but for the time-averaged intensity $I=4\times 10^{20}$~W/cm$^2$.} \label{fig:spectra4e20}
\end{figure*}

In Fig.~\ref{fig:spectra4e20}, we present the same as in Fig.~\ref{fig:spectra2e20} but for $I=4\times 10^{20}$~W/cm$^2$ and kinetic energies ranging from $10$~keV up to $35$~keV. One can see that, for 
these parameters, supercontinua are also formed. This time the maxima are located at $20.78$~keV for $\theta_{\bm p}=0.5\pi$ and $21.83$~keV for $\theta_{\bm p}=0.48\pi$. 
Therefore, we conclude that the position of the maximum in the energy spectra of photoelectrons scales linearly with the averaged intensity of the laser pulse (see, the results 
corresponding to $I=10^{20}$~W/cm$^2$ presented in Ref.~\cite{no_interference}). Note that just one saddle point contributes the most to the probability amplitude of ionization 
and $\textrm{Im}\,G(\phi_s)$, related to this particular saddle point, has a minimum around the region where the distribution is maximum (magenta curves in the lower panels). Moreover, comparing Figs.~\ref{fig:spectra2e20} and \ref{fig:spectra4e20}, one can see a considerable reduction of the maximum probability distribution (from one to two orders of magnitude, depending on the polar angle), which can be attributed to stabilization against ionization (see, e.g., Refs. \cite{professor,offerhaus,fedorov,gavrila,reiss2,muller,popov2}). 

\begin{figure*}
\begin{center}
\includegraphics[width=10.7cm]{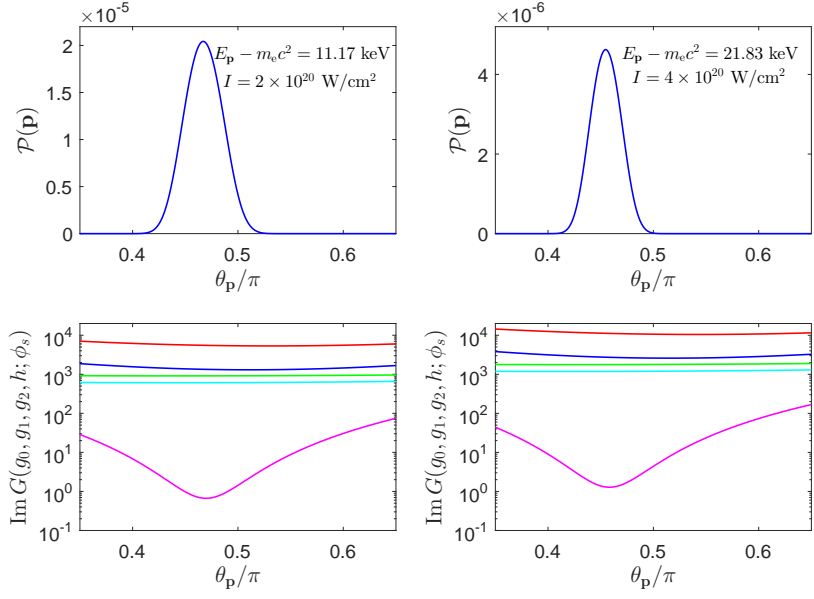}
\end{center}
\caption{(Color online) Initial-spin-averaged polar-angle distributions (upper row) for $\varphi_{\bm{p}}=0$ and for two time-averaged intensities: $I=2\times 10^{20}$~W/cm$^2$ (left column, maximum for $\theta_{\bm{p}}=0.467\pi$) and $I=4\times 10^{20}$~W/cm$^2$ (right column, maximum for $\theta_{\bm{p}}=0.455\pi$) with energies indicated in the figure. The maximum of these distributions scales as the inverse of intensity squared. In the lower row, we present the plots corresponding to $\textrm{Im}\,G(\phi_s)$ for all five relevant saddle points.} \label{fig:angular_spectra}
\end{figure*}

\begin{figure*}
\begin{center}
\includegraphics[width=12.9cm]{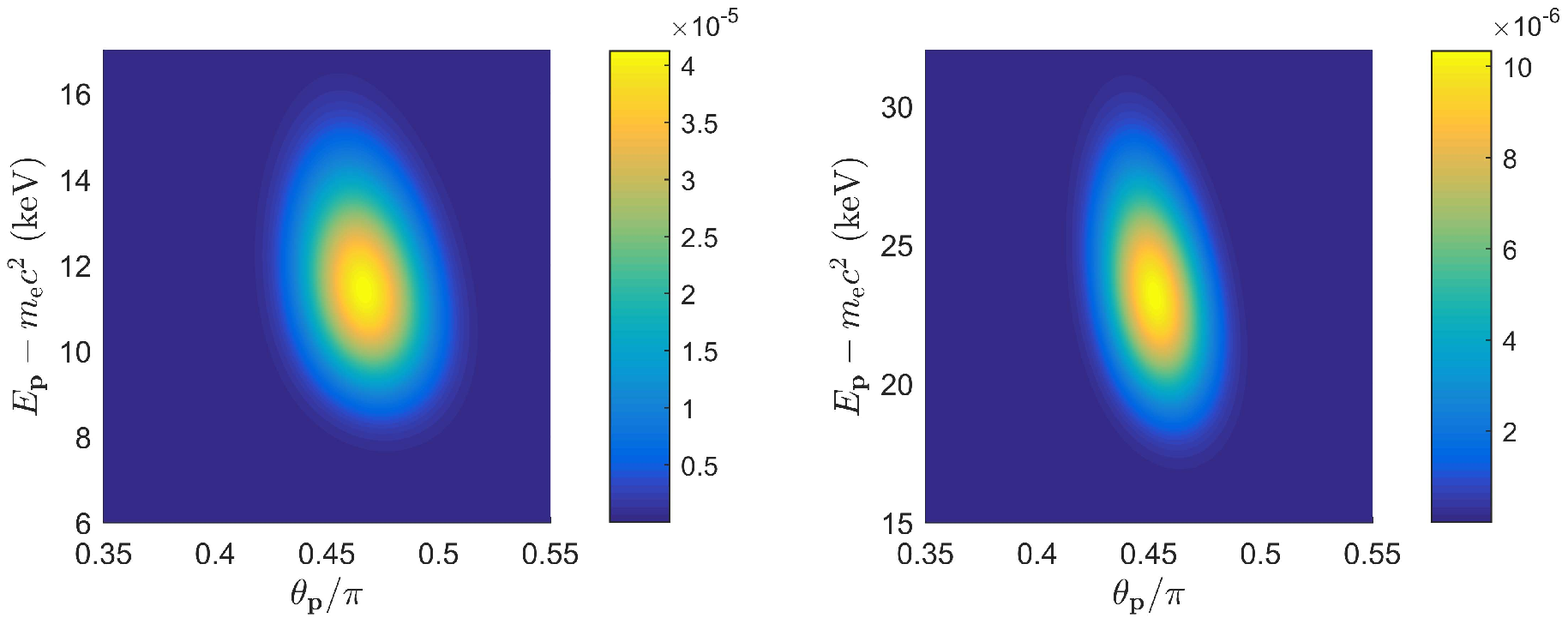}
\end{center}
\caption{(Color online) Color map of the probability distribution for the fixed azimuthal angle $\varphi_{\bm{p}}=0$ and for the time-averaged laser intensities $I=2\times 10^{20}$~W/cm$^2$ 
(left panel) and $I=4\times 10^{20}$~W/cm$^2$ (right panel). With increasing the field intensity the maximum of the distribution is shifted towards smaller polar angles $\theta_{\bm{p}}$ 
(radiation pressure effect) and towards higher energies (proportionally to the laser pulse intensity or the ponderomotive energy).
}\label{fig:energy_angular_probability}
\end{figure*}

In Fig.~\ref{fig:angular_spectra} the polar-angle spectra of photoelectrons at fixed kinetic energy are presented. The constant azimuthal angle has been chosen to be $\varphi_{\bm{p}}=0$ and the averaged intensities are $I=2\times 10^{20}$~W/cm$^2$ and $I=4\times 10^{20}$~W/cm$^2$ (upper left and right panels, respectively). One can see that, in both cases, a broad structure is formed. Additionally, as it was discussed in Refs.~\cite{no_interference,reiss2,reiss1}, the non-relativistic SFA predicts a maximum value of the distribution at $\theta_{\bm{p}}=0.5\pi$ for circularly polarized laser fields, which is not the case in our numerical calculations. One can see from the upper left panel in Fig.~\ref{fig:angular_spectra} that the actual maximum is located at $\theta_{\bm{p}}=0.467\pi$, i.e., it is shifted towards the direction of propagation of the laser field. Moreover, when the intensity is increased (upper right panel), the shifting is more pronounced and the maximum appears at $\theta_{\bm{p}}=0.455\pi$. Such effect has been attributed to the radiation pressure exerted by the laser field on the emitted photoelectrons \cite{no_interference,KMK2013,reiss2,reiss1}. On the other hand, one can clearly see that the maximum value of the distribution depends on the laser field intensity. More precisely, the maximum scales as the inverse of the averaged intensity squared, which is another indication of stabilization against ionization. In the lower panels of Fig.~\ref{fig:angular_spectra} we present the plots of $\textrm{Im}\,G(\phi_s)$ for all relevant saddle points. As expected, just one of them contributes importantly to the probability amplitude of ionization (magenta curves in the lower panels). In both cases, the minimum of $\textrm{Im}\,G(\phi_s)$ appears near the angular regions for which the distribution acquires maximum.

In Fig.~\ref{fig:energy_angular_probability} we present the color maps of the spectra of photoelectrons [Eq.~\eqref{dirac44}] as a function of the electron kinetic energy and polar angle for $\varphi_{\bm{p}}=0$. The time-averaged intensities are $I=2\times 10^{20}$~W/cm$^2$ (left panel) and $I=4\times 10^{20}$~W/cm$^2$ (right panel). One can see that, while the intensity increases, the position of the maximum of the distribution is shifted towards smaller polar angles and larger kinetic energies. The former is a consequence of radiation pressure, as it was discussed before. Furthermore, according to our calculations, the differential probability integrated over kinetic energy and polar angle decreases with increasing the laser field intensity, which is a signature of  stabilization against ionization.

\section{Conclusions}
We have analyzed the probability distribution of photoelectrons obtained from the interaction of relativistically-intense and short laser pulses with hydrogen-like ions under the RSFA framework. Our treatment is applicable even to very light ions, as it was illustrated for He$^+$, due to the fact that the ground-state wavefunction is well-defined before the interaction with the pulse. Furthermore, we have demonstrated that, by adjusting the parameters of the driving laser field, the energy spectrum of photoelectrons can exhibit a supercontinuum. Using the saddle-point approximation, we have related such broad structure to energy regions without interference (i.e., regions for which just one saddle point contributes importantly to the probability amplitude of ionization). Contrary to the results presented in Ref.~\cite{no_interference}, we have considered the initial-spin-averaged probability distributions, without restricting ourselves to fixed initial and final spin states. 

In our numerical calculations we have shown that the position of the maximum of the energy spectra of photoelectrons increases linearly with the averaged intensity of the driving field. Furthermore, the differential probability distribution integrated over kinetic energy and polar angle, for $\varphi_{\bm{p}}=0$, decreases with intensity, which can be attributed to stabilization against ionization. The polar-angle distribution presents a maximum at $\theta_{\bm p}<0.5\pi$ due to the radiation pressure exerted by the laser field. Moreover, an increase of the averaged intensity of the pulse leads to a maximum located at smaller polar angles.

\smallskip

\end{document}